\documentclass[11pt,a4paper,oneside,final]{article}

\pdfoutput=1
\usepackage{epsf,amsmath,amssymb,dcolumn}
\usepackage{scalefnt,cite,hyperref}
\usepackage{cleveref,etoolbox,color,soul}
\usepackage{float}
\usepackage{listings}
\usepackage[utf8]{inputenc}
\usepackage{indentfirst}

\usepackage{caption}
\usepackage{subcaption,graphicx}

\Crefname{figure}{Fig.}{Figs.}
\usepackage{placeins}
\usepackage{soul}

\usepackage[normalem]{ulem}

\usepackage[utf8]{inputenc}
\hypersetup{
	unicode,
	pdfproducer={LaTeX},
	pdfcreator={pdflatex}
}

\newcounter{notecount}
\begin{document}
\thispagestyle{empty}

\long\def\symbolfootnote[#1]#2{\begingroup%
\def\thefootnote{\fnsymbol{footnote}}\footnote[#1]{#2}\endgroup}

\vspace{1cm}

\begin{center}
\Large\bf\boldmath
Silicon-Photomultiplier (SiPM) Protection Against Over-Current and Over-Illumination
\unboldmath
\end{center}
\vspace{0.05cm}

\begin{center}
Gholamreza Fardipour Raki$^a$\footnote{Corresponding Author, fardipour@ipm.ir},
Mohsen Khakzad$^a$\footnote{mohsen@ipm.ir}\\[0.4em]
\end{center}

\begin{center}
{\small
{{\sl ${}^a$ School of particles and accelerator, Institute for Research in fundamental sciences (IPM), Tehran, Iran}}\\[0.2em]
}
\end{center}

\vspace*{25mm}
\begin{abstract}
\noindent

SiPMs operate in Geiger mode, wherein photodiode cells are reverse-biased to the breakdown by even a single photon. Each cell is connected in series with a quenching resistor, which prevents cell damage and resets the cell after making a signal. All cells are arranged in parallel, making SiPMs and biasing circuits vulnerable to over-illumination, where the current passing through the SiPM can exceed the allowable value, leading to damage. In this study, we investigate over-current conditions in SiPMs and propose a protective method against over-illumination and over-current using a series resistor. Additionally, we ensure SiPM stability through the incorporation of a suitable capacitor. 

\end{abstract}
\vspace*{30mm}


\setcounter{footnote}{0}

\newpage

\pagenumbering{arabic}

\section{Introduction to SiPM structure and the current and the illumination limits}
\label{sec:introduction}

\begin{figure}[H]
\centering
\begin{subfigure}{0.4\textwidth}
\includegraphics[width=1\textwidth]{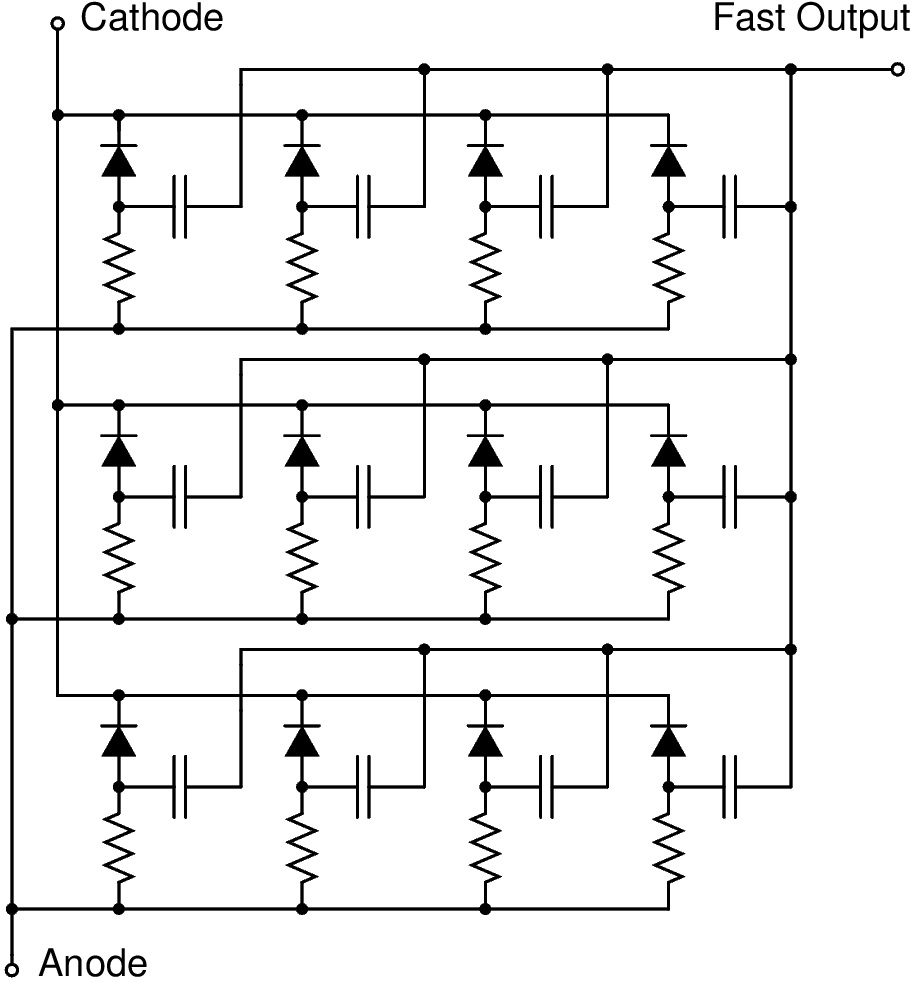} 
\end{subfigure}
\hspace*{2mm}
\begin{subfigure}{0.4\textwidth}
\includegraphics[width=1\textwidth]{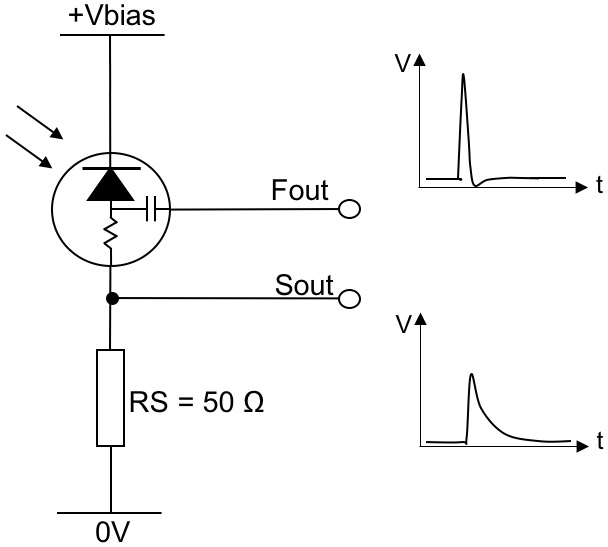}
\end{subfigure}
\vspace*{1mm}
\caption{(Left) Schematic of the SiPM structure. (Right) Biasing and readout circuits for C-Series ON-Semiconductors SiPM.}
\label{fig:SiPM}
\end{figure}

The Silicon-Photomultiplier (SiPM) is composed of an array of cells, with each cell containing a Single Photon Avalanche Diode (SPAD) and a quench resistor, $R_Q$. In SiPM driver circuits, the sense resistor, $R_S$, establishes a voltage output on the Standard output (Sout). To minimize the cell recovery time, the series resistance can be set to $R_S$ = 0. However, in this scenario, only the fast output is available. Figure~\ref{fig:SiPM} (left) illustrates the schematic of the SiPM structure, while Figure~\ref{fig:SiPM} (right) depicts the biasing and readout circuits for the C-Series ON-Semiconductors SiPM \cite{semiconductorc, semiconductorbiasing}.

\begin{figure}[H]
\centering
\begin{subfigure}{0.5\textwidth}
\includegraphics[width=1\textwidth]{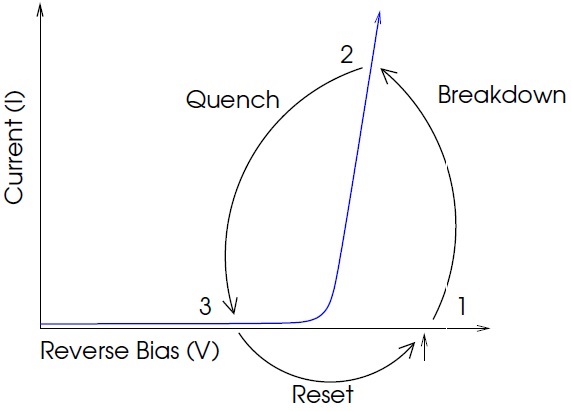}
\end{subfigure}
\vspace*{1mm}
\caption{Cycle of breakdown, quench, and reset in a SiPM operating in Geiger mode.}
\label{fig:avalanche}
\end{figure}

A photo-diode operating in Geiger mode utilizes the breakdown mechanism to get high gain and is known as a SPAD. By applying a reverse bias higher than its nominal breakdown voltage, high gradients of electric field are created across the junction, facilitating the generation of avalanches. Once a current begins flowing, it needs to be promptly halted or 'quenched'. Quenching is achieved through a series resistor, $R_Q$, which restricts the current flowing through the cell's photo-diode during breakdown. This lowers the reverse voltage across the photo-diode to a level below its breakdown voltage, effectively stopping the avalanche. Subsequently, the photo-diode recharges back to the main biasing voltage, becoming ready to detect new photons. The breakdown, avalanche, quench, and recharge cyclic process with returning of the biasing voltage to a value above the breakdown, is shown in Figure~\ref{fig:avalanche} \cite{AND9770/D, Raki:2022lwn}. 

The SiPM incorporates a dense array of tiny, independent SPAD detectors, each accompanied by a quenching resistor. Any independent SPAD and its corresponding quench resistor are typically mentioned to as a "cell". If a cell within the SiPM responds to a photon, it initiates a Geiger avalanche, resulting in the generation of a photo-current within the cell. As a consequence, there is a voltage drop by the quench resistor, diminishing the voltage bias across the photo-diode to an amount below the breakdown. This action effectively quenches the photo-current and halts further Geiger-mode avalanches. Following this, once the photocurrent is quenched, and the voltage across the photo-diode restores to its nominal bias value. The duration required for the cell to recharge to the full operating voltage is termed the recovery time. 

It is essential to emphasize that the Geiger avalanche remains localized to the individual cell where it originated. While the avalanche occurs in one cell, all other cells remain fully charged and prepared to detect photons. Given that SiPMs are highly sensitive photodetectors, capable of detecting very low levels of signal light, precautions must be taken to shield against extraneous light sources \cite{semiconductorc, semiconductorbiasing}. Quenching resistors typically range between 150 k$\Omega$ to 1 M$\Omega$ \cite{renker2009advances, jiang2017recovery}.

We utilized the C-Series SiPM from ON-Semiconductor, specifically the MICROFC-60035-SMT-TR1 model, for our study. This device comprises 60,035 micro-cells arranged in parallel. The breakdown voltage ($V_{br}$) for this device typically falls within the range of 24.2 V to 24.7 V. The recommended overvoltage range (voltage above $V_{br}$) is typically advised to be between 1 V to 5 V. It is crucial to consider the maximum current level, which is specified as 20 mA when applying the overvoltage \cite{semiconductorc}. With the lowest biasing voltage set at 25 V, in order to ensure that the electric current remains below the maximum allowed value when the SiPM is exposed to high-intensity light, the effective resistance of the SiPM cannot be less than 1250 $\Omega$.

During exposure to intense light, the cells operate in Geiger mode, undergoing the cycle of breakdown, quenching, and reset. During breakdown, the resistance of the cell is nearly equivalent to the quench resistance. Considering the SiPM with the highest quench resistance, set at 1 M$\Omega$, only approximately 800 cells are permitted to be in the breakdown stage simultaneously. This constitutes around 1 percent of the total number of cells. Therefore, it is evident that the current in the SiPM can easily reach the maximum allowed value if not adequately shielded against environmental light. Overcurrent has the potential to damage the SiPM and may even affect the power supply.

\section{SiPM protection method against the over-current}
\label{sec:protection}

\begin{figure}[H]
\centering
\begin{subfigure}{0.9\textwidth}
\includegraphics[width=1\textwidth]{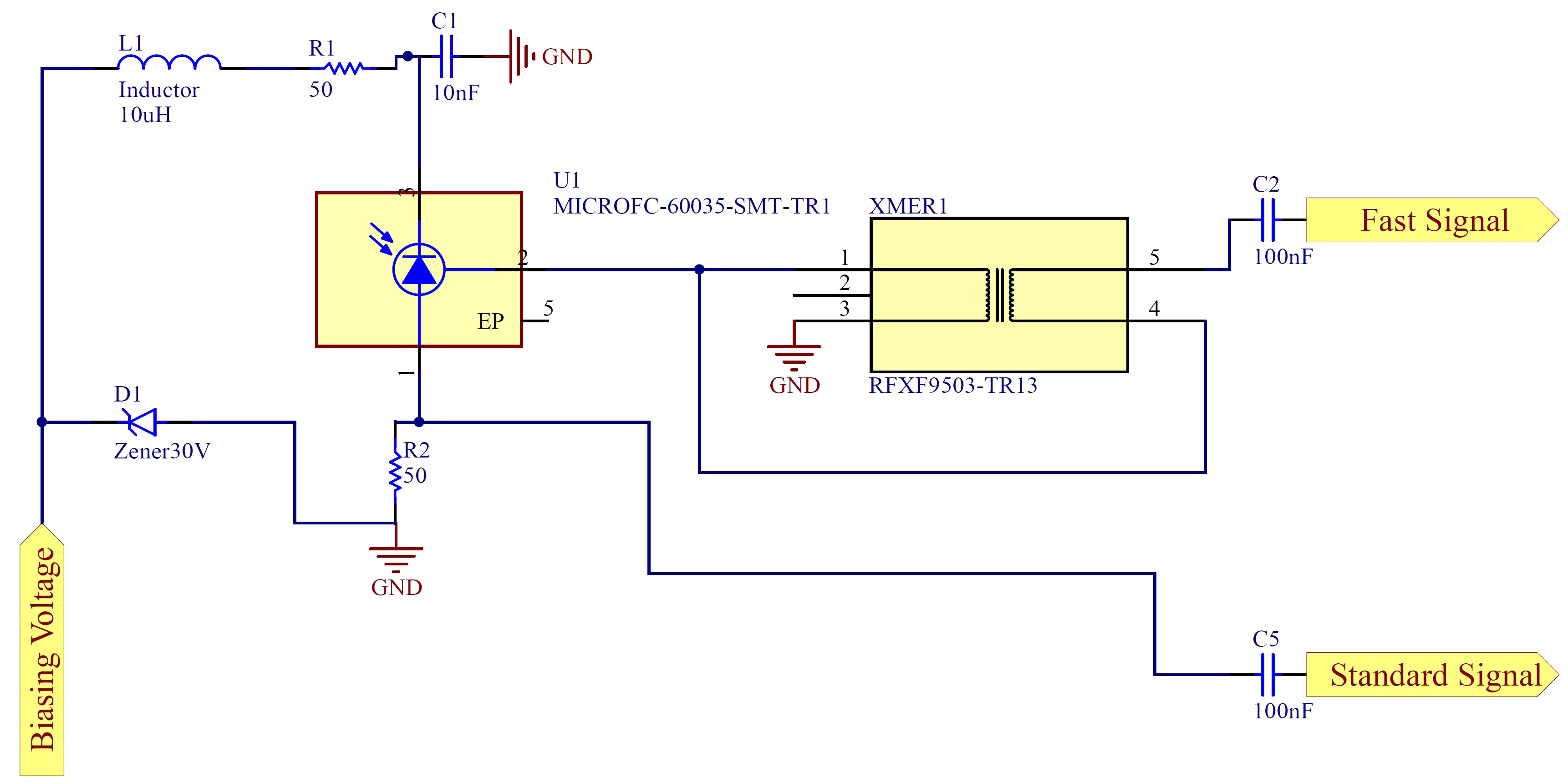} 
\end{subfigure}
\hspace*{2mm}
\begin{subfigure}{0.7\textwidth}
\includegraphics[width=1\textwidth]{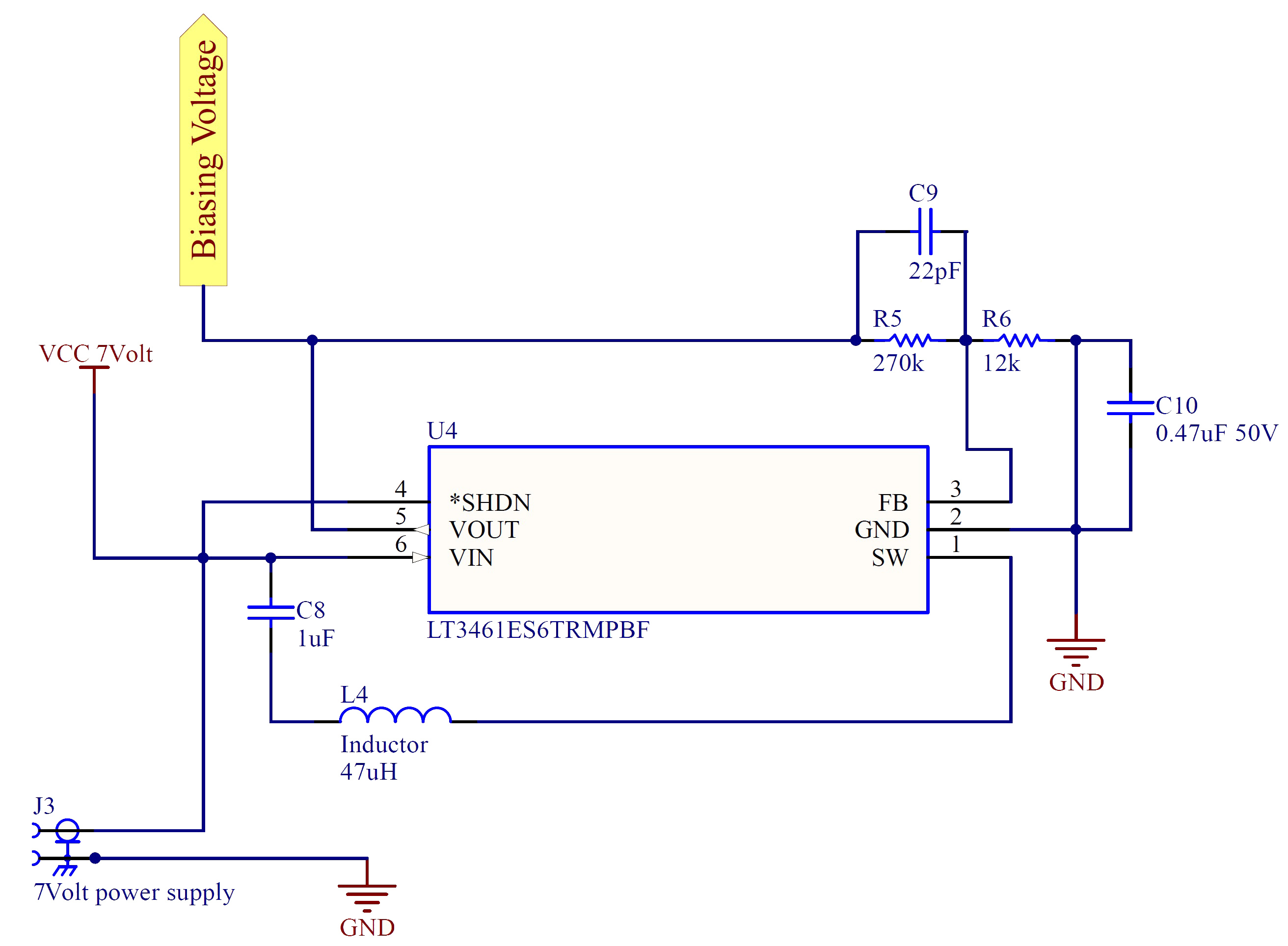}
\end{subfigure}
\vspace*{1mm}
\caption{(top) The schematic of the SiPM driver. (bottom) The schematic of the SiPM biasing circuit.}
\label{fig:driver}
\end{figure}

Indeed, shielding against extraneous light is crucial to prevent over-illumination of SiPMs. Additionally, safeguarding the biasing circuit from over-voltage is essential for maintaining the integrity of SiPM operation and preventing damage. One effective method for achieving this is by incorporating a zener diode into the circuitry. We have developed a standard driver circuit for SiPMs to ensure reliable and stable operation, as depicted in Figure~\ref{fig:driver} \cite{Raki:2022lwn}. This circuit design includes provisions for voltage regulation and protection against over-voltage conditions, thereby enhancing the robustness of SiPM performance in diverse applications.

\begin{figure}[H]
\centering
\begin{subfigure}{0.7\textwidth}
\includegraphics[width=1\textwidth]{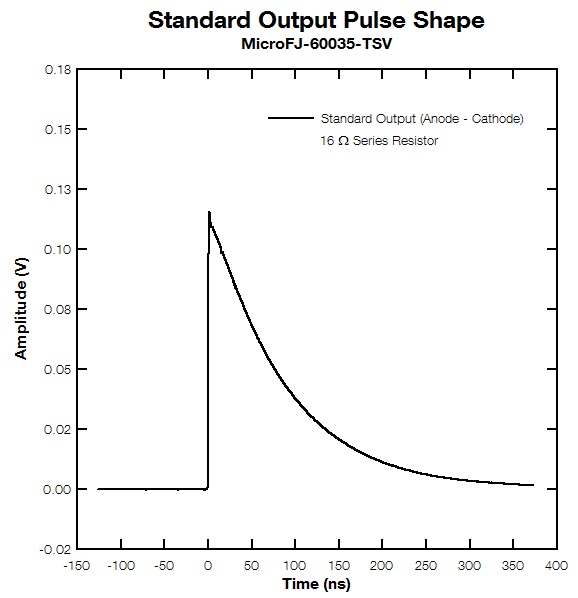}
\end{subfigure}
\vspace*{1mm}
\caption{Standard output pulse shape of an ON-Semiconductor SiPM, microFJ-60035-TSV}
\label{fig:signal}
\end{figure}

The maximum current for ON-Semiconductor products, and similar ranges for others, can be between 6 mA to 20 mA, thus the allowed power dissipation in the studied SiPM is at most 0.6 W. Considering the typically very low signal light levels to be detected in the SiPM, along with accounting for dark current and possible background signals \cite{renker2009advances, raki2023cosmic}, it remains much lower than 0.6 W. Figure~\ref{fig:signal} \cite{AND9770/D} shows a common standard signal from ON-Semiconductor SiPM, microFJ-60035-TSV, with a signal duration in the range of 100 ns.

\begin{figure}[H]
\centering
\begin{subfigure}{0.7\textwidth}
\includegraphics[width=1\textwidth]{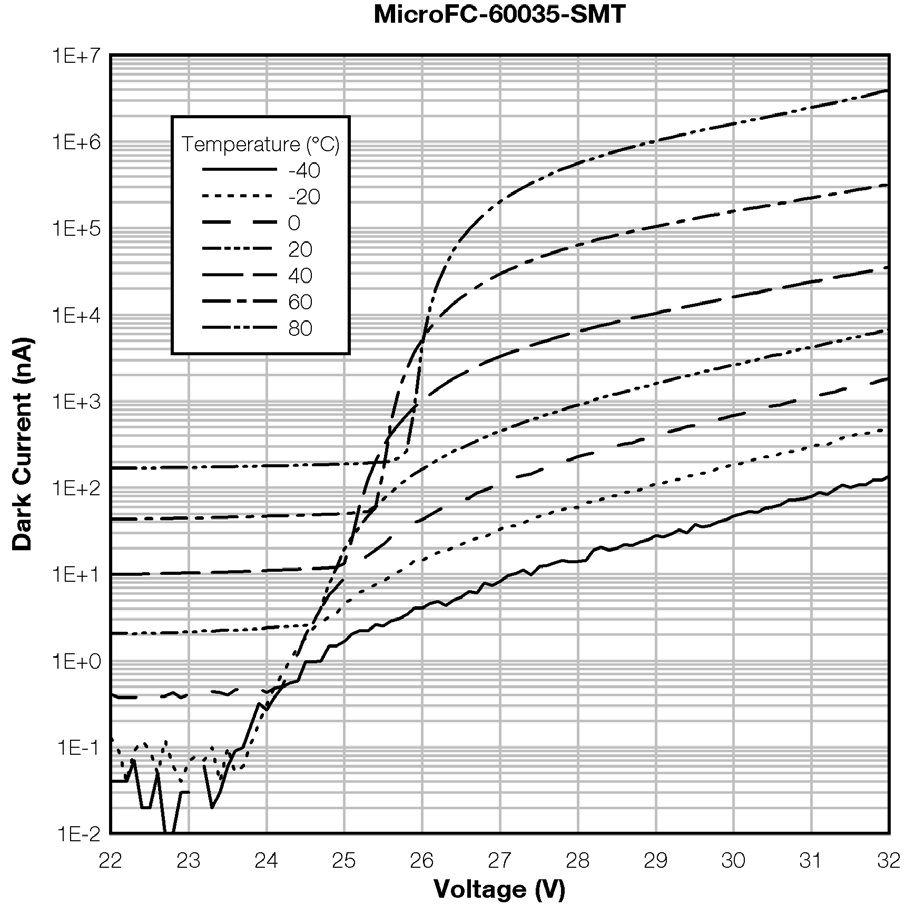}
\end{subfigure}
\vspace*{1mm}
\caption{Dark current versus biasing voltage and temperature.}
\label{fig:dark}
\end{figure}

Figure~\ref{fig:dark} illustrates the dark current versus reverse biasing voltage and temperature for the ON-Semiconductor SiPM, microFC-60035-SMT \cite{semiconductorc}. At room temperature, the dark current for a biasing voltage of 30 V is less than 10$\mu$A, resulting in a power dissipation of approximately 0.3 mW.

\begin{figure}[H]
\centering
\begin{subfigure}{0.5\textwidth}
\includegraphics[width=1\textwidth]{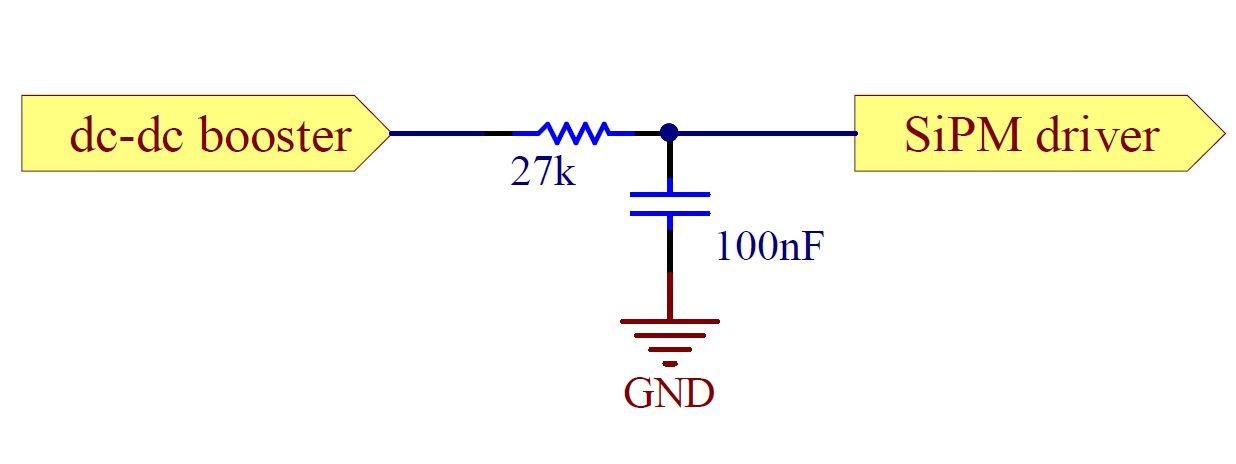}
\end{subfigure}
\vspace*{1mm}
\caption{Protection circuit between biasing and SiPM driver circuit.}
\label{fig:protect}
\end{figure}

As shown in Figure~\ref{fig:protect}, the main component, a 27 k$\Omega$ resistor, is suggested to be added to the SiPM circuit (Figure~\ref{fig:driver}) just after the biasing circuit. The voltage drop in the 27 k$\Omega$ resistor due to the dark current (10 $\mu$A) is 0.27 V and this can be adjusted to compensate easily in the DC-DC booster by varying the resistor $R_5$ (Figure~\ref{fig:driver} bottom) to an amount a bit higher than 270 k$\Omega$ \cite{Raki:2022lwn}.
In the event of over-current or over-illumination, if the current reaches 1/20 of the maximum allowed, equal to 1 mA, the voltage drop will be 27 V, making it impossible to reach a dangerous current level with the addition of the 27k$\Omega$ resistor. Even in an imaginary scenario where all the cells in the SiPM are in breakdown simultaneously, the SiPM resistance cannot be less than 16 ohms, the 27 k$\Omega$ series resistor causes a voltage drop that effectively protects the SiPM and DC-DC booster against damage.

As depicted in Figure~\ref{fig:signal}, the standard signal length is approximately 100 ns. Thus, if we were even extend this duration by 10 times, while maintaining the biasing voltage at 30 V with the maximum allowed current, denoted as $U$, the required energy can be calculated as follows:

\begin{equation}
\label{eq:energy}
U= VIt = 30(V) \times20 (A) \times10^{-3}\times1000 (s) \times10^{-9} = 600\times10^{-9} J 
\end{equation}

The second major component suggested to add to the SiPM driver is a 100 nF ceramic capacitor connected in parallel to the entrance of the biasing voltage, parallel to the Zener diode $D_1$ in Figure~\ref{fig:driver} (top). The energy, denoted as $u$, saved by the 30 V biasing voltage in this capacitor can be calculated using the formula:

\begin{equation}
\label{eq:capenergy}
u= \frac{1}{2} CV^2 = \frac{1}{2}\times100 (F) \times10^{-9}\times900 (V^2)= 45000\times10^{-9} J 
\end{equation}

Thus, the energy saved in the capacitor is sufficient to handle the SiPM in the event of receiving light signals. The time taken for the charge on the capacitor to reach 63 percent of its maximum possible fully charged state, known as the Time Constant ($T$), can be calculated as follows:

\begin{equation}
\label{eq:time}
T= RC = 27 (\Omega) \times10^{3}\times100 (F) \times10^{-9} = 2.7\:ms 
\end{equation}

Therefore, in each second, it is possible to charge the capacitor up to 370 times.

\section{Conclution}
\label{sec:conclution}

Therefore, with the method outlined above, it is feasible to drive the SiPM and acquire light signals while ensuring protection against over-current and over-illumination. The circuit and method were tested, and the optimized values for the resistor (27 k$\Omega$) and capacitor (100 nF) were selected. The protection circuit was evaluated for its efficacy in capturing signals from cosmic muons \cite{raki2023cosmic}, detected by a plastic scintillator BC-480 (5 cm $\times$ 15 cm $\times$ 15 cm). The addition of the protection circuit did not alter the signal count or shape. Moreover, when the SiPM was exposed to direct sunlight for approximately 1 hour without shielding, its operation almost ceased. However, normal functionality was restored after shielding was applied.

\section*{Acknowledgments}
%
We are especially grateful to the School of Particles and Accelerators at IPM for the continual interest shown in our projects, and the financial support they provided.

%
%



{
\bibliographystyle{utphys}
\bibliography{reference.bib}
}

\end{document}